\documentclass[aip,reprint,author-year,nofootinbib]{revtex4-1}

\usepackage{graphicx}
\usepackage{amsmath,amssymb}             
\usepackage{stmaryrd}					
\usepackage{color}
\usepackage[dvipsnames]{xcolor}
\usepackage{color,hyperref}
\definecolor{darkblue}{rgb}{0.0,0.0,0.4}
\definecolor{red}{rgb}{0.7,0.0,0.0}
\definecolor{green}{rgb}{0.0,0.5,0.0}
\hypersetup{colorlinks,breaklinks,
            linkcolor=darkblue,urlcolor=darkblue,
            anchorcolor=darkblue,citecolor=RoyalBlue}

\def\apj{Astrophys. J.}

\def\apjs{Astrophys. J. Supp.}

\def\mnras{Mon. Not. R. Astron. Soc.}

\def\aap{Astron. \& Astrophys.}

\def\jcap{Journal of Cosmology and Astroparticle Physics}

\defcitealias{Jasche2015BORGSDSS}{\textsc{borg sdss}}
\newcommand{\borg}{\textsc{borg}}
\newcommand{\tweb}{\textsc{T-web}}
\newcommand{\diva}{\textsc{diva}}
\newcommand{\origami}{\textsc{origami}}

\begin{document}


\title{Probabilistic cartography of the large-scale structure}


\author{Florent Leclercq}
\email{florent.leclercq@polytechnique.org}
\affiliation{Institute of Cosmology and Gravitation, University of Portsmouth,\\ Dennis Sciama Building, Burnaby Road, Portsmouth, PO1 3FX, United Kingdom}
\affiliation{Institut d'Astrophysique de Paris (IAP), UMR 7095, CNRS -- UPMC Universit\'e Paris 6, Sorbonne Universit\'es, 98bis boulevard Arago, F-75014 Paris, France}
\affiliation{Institut Lagrange de Paris (ILP), Sorbonne Universit\'es,\\ 98bis boulevard Arago, F-75014 Paris, France}
\affiliation{\'Ecole polytechnique ParisTech,\\ Route de Saclay, F-91128 Palaiseau, France}

\author{Jens Jasche}
\affiliation{Excellence Cluster Universe, Technische Universit\"at M\"unchen,\\ Boltzmannstrasse 2, D-85748 Garching, Germany}

\author{Guilhem Lavaux}
\affiliation{Institut d'Astrophysique de Paris (IAP), UMR 7095, CNRS -- UPMC Universit\'e Paris 6, Sorbonne Universit\'es, 98bis boulevard Arago, F-75014 Paris, France}
\affiliation{Institut Lagrange de Paris (ILP), Sorbonne Universit\'es,\\ 98bis boulevard Arago, F-75014 Paris, France}

\author{Benjamin Wandelt}
\affiliation{Institut d'Astrophysique de Paris (IAP), UMR 7095, CNRS -- UPMC Universit\'e Paris 6, Sorbonne Universit\'es, 98bis boulevard Arago, F-75014 Paris, France}
\affiliation{Institut Lagrange de Paris (ILP), Sorbonne Universit\'es,\\ 98bis boulevard Arago, F-75014 Paris, France}
\affiliation{Department of Physics, University of Illinois at Urbana-Champaign,\\ 1110 West Green Street, Urbana, IL~61801, USA}
\affiliation{Department of Astronomy, University of Illinois at Urbana-Champaign,\\ 1002 West Green Street, Urbana, IL~61801, USA}


\date{\today}

\begin{abstract}
\noindent The {\borg} algorithm is an inference engine that derives the initial conditions given a cosmological model and galaxy survey data, and produces physical reconstructions of the underlying large-scale structure by assimilating the data into the model. We present the application of {\borg} to real galaxy catalogs and describe the primordial and late-time large-scale structure in the considered volumes. We then show how these results can be used for building various probabilistic maps of the large-scale structure, with rigorous propagation of uncertainties. In particular, we study dynamic cosmic web elements and secondary effects in the cosmic microwave background.
\end{abstract}


\maketitle



\section{Bayesian large-scale structure inference with BORG}

Over the last few years, several models and software packages aiming at full analysis of the three-dimensional cosmological matter distribution have met some success. Among them, {\borg} \citep[Bayesian Origin Reconstruction from Galaxies,][]{Jasche2013BORG} is a full-scale Bayesian inference framework for analyzing the linear and mildly non-linear large-scale structure.

Contrary to previous approaches, which relied on phenomenological density models, {\borg} involves an additional layer of complexity by running several numerical simulations of structure formation for each move in a huge parameter space, comprising of the order of $10^7$ parameters (the voxels of the discretized domain). In this fashion, the data model jointly accounts for the shape of three-dimensional matter field and its formation history. To allow feasible numerical analyses, {\borg} relies on the Hamiltonian Monte Carlo algorithm. The (approximate) physical model for gravitational dynamics is second-order Lagrangian perturbation theory (2LPT), linking initial density fields (at a scale factor $a=10^{-3}$) to the presently observed large-scale structure (at $a=1$). The galaxy distribution is modeled as an inhomogeneous Poisson process on top of evolved density fields. In its latest version, {\borg} also accounts for luminosity dependent galaxy biases and performs automatic calibration of corresponding noise levels \citep{Jasche2015BORGSDSS}. For a more extensive discussion of the {\borg} data model, the reader is referred to chapter 4 in \citet{LeclercqThesis}.

Over the last two years, Bayesian large-scale structure inference with {\borg} has moved beyond the proof-of-concept stage, to routine application to real data such as the Sloan Digital Sky Survey (SDSS) main galaxy sample \citep{SDSSDR7} and the 2M++ catalog \citep{Lavaux2MPP}: see \citet{Jasche2015BORGSDSS,Lavaux2016BORG2MPP}.

\begin{figure*}
\begin{center}
\includegraphics[width=\textwidth]{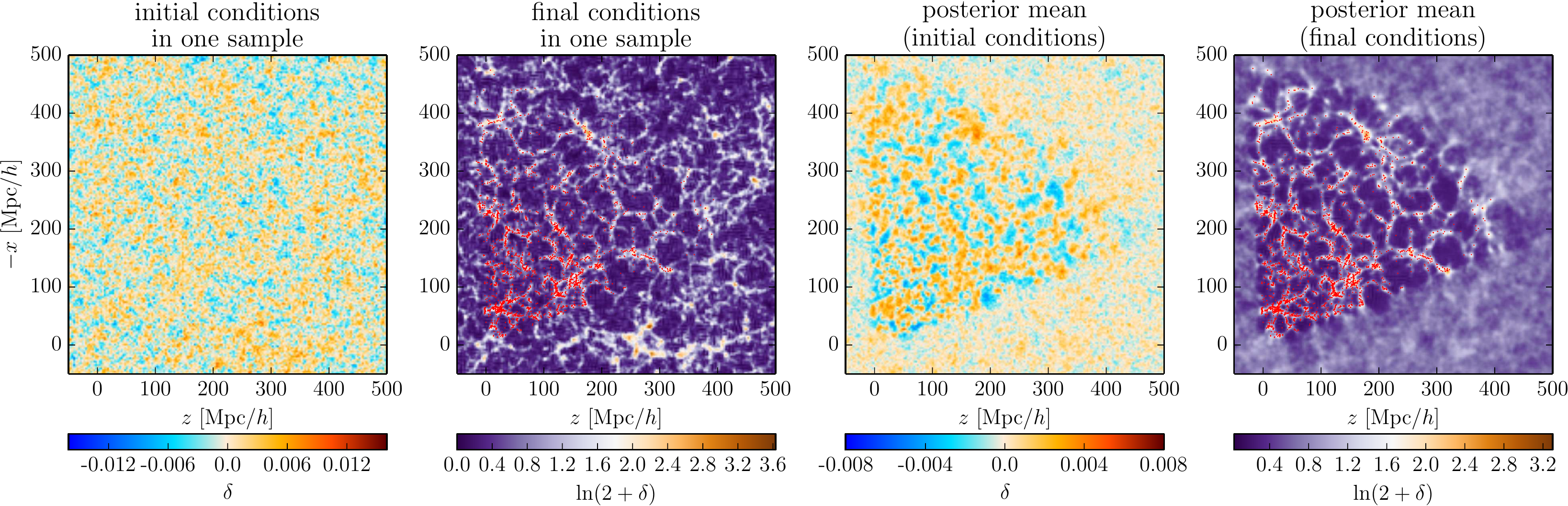}
\end{center}
\caption{Bayesian large-scale structure inference with {\borg} in the SDSS main galaxy sample. \textit{Leftmost panels}: slices through one sample of the posterior for the initial and final density fields. \textit{Rightmost panels}: posterior mean in the initial and final conditions. The input galaxies are overplotted on the final conditions as red dots.\label{fig:overview_galaxies}}
\end{figure*}

\begin{figure*}
\begin{center}
\includegraphics[width=\textwidth]{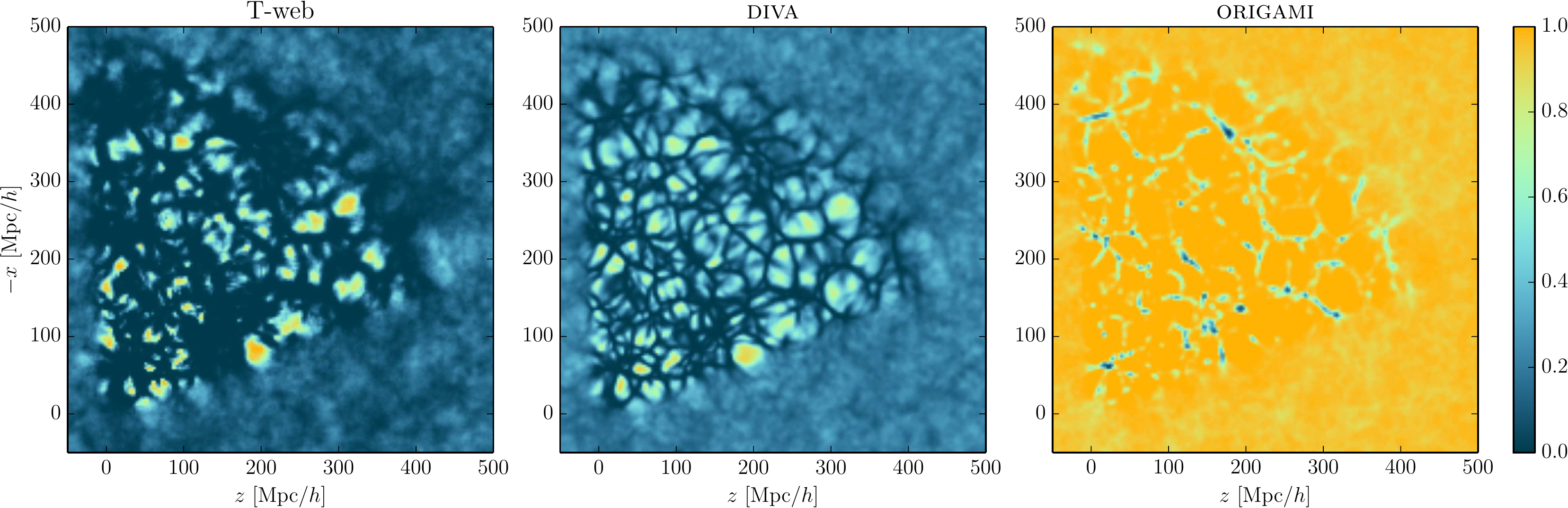}
\end{center}
\caption{Comparison of cosmic web classification procedures in the SDSS volume. The panels show slices through the posterior probability for voxels to belong to a void, as defined by the {\tweb} (left panel), by {\diva} (middle panel) and by {\origami} (right panel). \label{fig:comparison_pdf_voids}}
\end{figure*}

In figure \ref{fig:overview_galaxies}, we illustrate the results of Bayesian large-scale structure inference with {\borg}. The two leftmost panels show slices through the reconstructed density in one sample, in the initial conditions (at $a=10^{-3}$) and in the corresponding final conditions (at $a=1$). The SDSS galaxies are overplotted as red dots. In our Bayesian framework, each of the constrained samples is a full-scale realization of the physical model (2LPT), and the variation between samples quantifies uncertainties. In the two rightmost panels, we show the ensemble mean among all the samples obtained in our analysis, which approximates the posterior mean, for initial and final conditions. The mean density field exhibits a high degree of detail where data constraints are available, but approaches cosmic mean density in unobserved regions (at high redshift or out of the survey boundaries).

Beyond density reconstruction, Bayesian large-scale structure inference with {\borg} yields a rich variety of scientific products. In the following, we illustrate with two examples: cosmic web classification and production of templates for secondary effects expected in the cosmic microwave background (CMB). A particular advantage of this approach is that it automatically and self-consistently propagates observational uncertainties from the inferred density to other physical quantities.

\begin{figure*}
\begin{center}
\begin{tabular}{cc}
\footnotesize{iSW} & \footnotesize{iSWRS} \\[-25pt]
\includegraphics[width=0.45\textwidth]{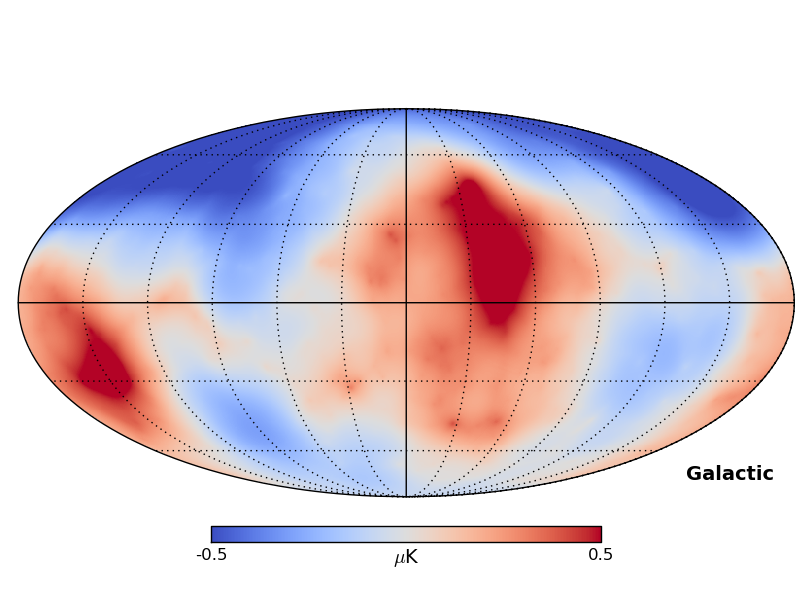} & \includegraphics[width=0.45\textwidth]{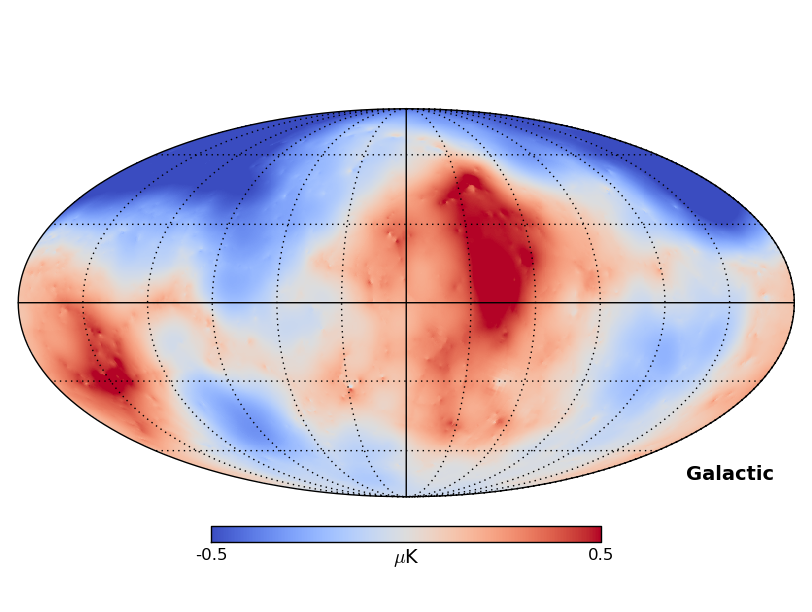} \\[5pt]
\footnotesize{RS} & \footnotesize{kSZ} \\[-25pt]
\vspace{-10pt}
\includegraphics[width=0.45\textwidth]{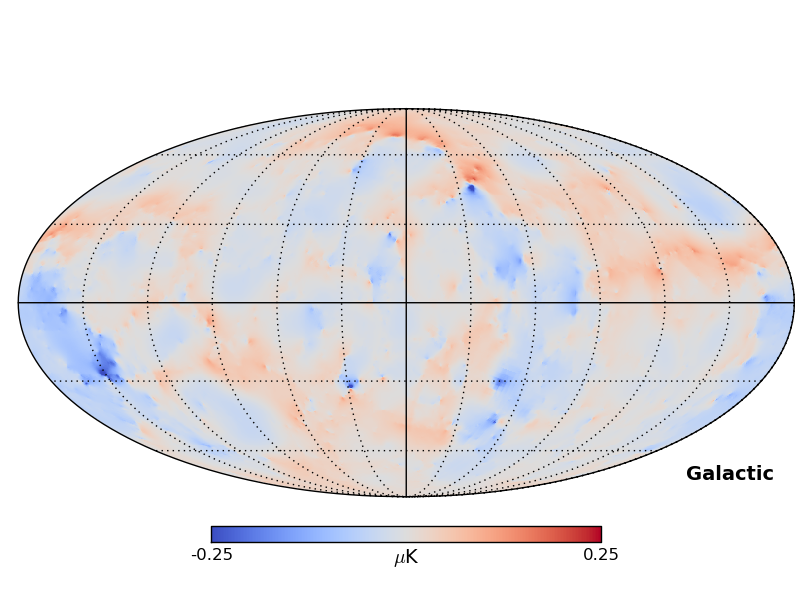} & \includegraphics[width=0.45\textwidth]{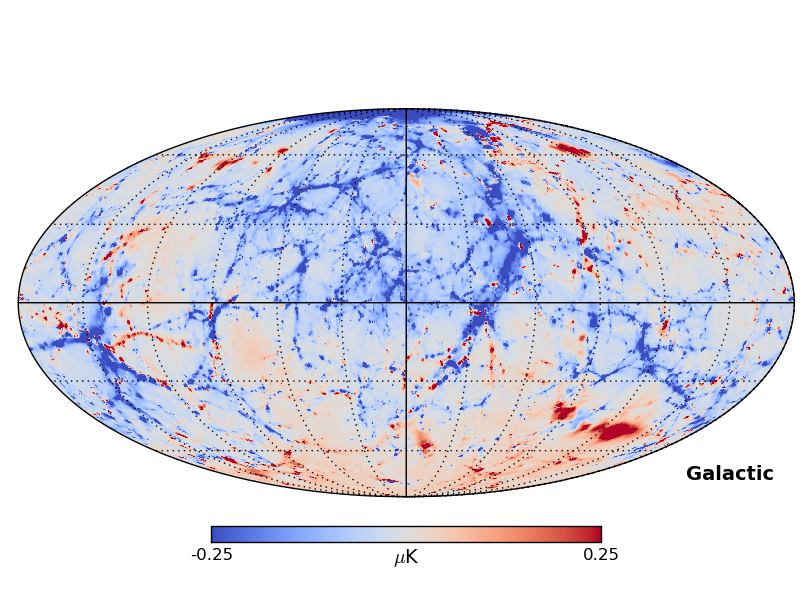}
\end{tabular}
\end{center}
\caption{Templates for CMB secondary effects. The four panels correspond to, respectively: iSW (using the linear gravitational potential), iSWRS (using the fully non-linear gravitational potential), RS (obtained by subtracting the previous maps, i.e. iSWRS$-$iSW) and kSZ.\label{fig:cmb_templates}}
\end{figure*}

\section{Cosmic web classification}

As demonstrated in \citet{Leclercq2015ST} for the SDSS, {\borg} inference results can be used as inputs for a detailed cosmic-web type analysis. The large-scale structure is dissected and classified in terms voids, sheets, filaments, and clusters. The resulting cosmic web maps are fully probabilistic: in each voxel, four probabilities (summing up to unity) for each of the structure types are obtained.

In \citet{Leclercq2015ST}, the classification procedure adopted is the so-called {\tweb} algorithm \citep{Hahn2007}, which consists in looking at the eigenvalues of the tidal tensor field. The number of positive (resp. negative) eigenvalues corresponds to the number of axes along which gravitational collapse (resp. expansion) occurs, which naturally classifies the environment into clusters, sheets, filaments and voids. As the tidal tensor is directly derived from the density field, the {\tweb} can be applied to final conditions reconstructed with any data model \citetext{see \citealp{Jasche2010}, for an earlier application of the {\tweb} to density fields reconstructed using a log-normal density model}. However, {\borg} allows a chrono-cosmographic description of the dynamic cosmic web, in the sense that it also infers proto-structures present in the initial conditions and their time evolution.

Further, the inference of the initial density field by {\borg} now allows a description of the cosmic web in real data using ``Lagrangian classifiers'' \citep{LeclercqLAGRANGIAN}, i.e. algorithms that necessitate the initial positions of particles. Among them, {\diva} \citep{LavauxDIVA} uses the eigenvalues of the Lagrangian displacement field, and {\origami} \citep{Falck2012} counts the number of shell-crossings. This new possibility offered by physical large-scale structure inference is of special interest, because the use of Lagrangian classifiers has so far been limited to simulations. 

In figure \ref{fig:comparison_pdf_voids}, we compare probabilistic maps of the SDSS volume for voids, as defined by the {\tweb} \citep[left panel, reproduced from][]{Leclercq2015ST}, by {\diva} \citep[middle panel, reproduced from][]{LeclercqLAGRANGIAN} and by {\origami} \citep[right panel, reproduced from][]{LeclercqLAGRANGIAN}. The {\tweb} and {\diva} maps are visually similar, with an overall smoother structure for the voids defined by {\diva}, which are sharply separated by sheets and filaments. In contrast, with {\origami}, most of the volume is filled by voids (this is also true for the prior) and more complex, shell-crossed structures are rarely identified.

These developments naturally bring in a connection between cosmic web analysis and information theory. In \cite{Leclercq2015ST}, we examine the Shannon entropy of the structure-type posterior probability distribution and quantify the information gain due to SDSS galaxies. In \cite{Leclercq2015DT}, we propose a decision criterion for classifying structures in the presence of uncertainty. The resulting decision-making procedure balances the posterior probabilities and the strength of data constraints. Finally, in~\cite{LeclercqCOMPARISON}, we extend the problem to the space of classifiers, and introduce utility functions for the optimal choice of a classifier, specific to the application of interest.

\section{Secondary effects in the cosmic microwave background}

Beyond analyses of the large-scale structure as probed by galaxies, {\borg} inference results can be used to produce templates for secondary effects expected in the CMB: the kinetic Sunyaev-Zel'dovich (kSZ) effect, the integrated Sachs-Wolfe (iSW) and Rees-Sciama (RS) effects. The cross-correlation of such templates with CMB maps, for example via a matched-filter approach~\citep{Li2014}, can then enhance the detectability of these effects. 

Starting from initial conditions produced by the {\borg} 2M++ analysis \citep{Lavaux2016BORG2MPP}, we generate a large ensemble of constrained simulations using the fast \textsc{cola} model \citep{Tassev2013}. These simulations describe complex non-linear dynamics in the nearby Universe, which imprints effects on CMB photons: the momentum field of electrons (approximated to the momentum field of the matter field), resulting in the kSZ effect; and the time evolution of the gravitational potential (linear and non-linear), resulting in the iSW and iSWRS effects. We use the kSZ data model presented in~\citet{LavauxkSZ}, and the iSW/iSWRS models as well as the ray-tracing algorithm presented in~\citet{Cai2010}. In \citet{LavauxCMB}, we present the resulting templates and show that better physical modeling, as made possible by {\borg}, yields higher signal-to-noise ratio when analyzing CMB secondary effects.

Figure \ref{fig:cmb_templates} shows examples of templates, produced using raytracing from $0$ to $100$~Mpc/$h$. Only one sample is shown here, but as before, the full Bayesian posterior is available for thorough quantification of uncertainties.

\acknowledgments

\small{FL thanks the organizers of the Rencontres du Vietnam 2015, \textit{Cosmology 50 years after CMB discovery}, for a very nice meeting and acknowledges support from the \'Ecole polytechnique through an AMX grant and from the European Research Council through grant 614030, Darksurvey. JJ is partially supported by a Feodor Lynen Fellowship by the Alexander von Humboldt foundation. BW acknowledges funding from an ANR Chaire d'Excellence (ANR-10-CEXC-004-01) and the UPMC Chaire Internationale in Theoretical Cosmology. This work has been done within the Labex \href{http://ilp.upmc.fr/}{Institut Lagrange de Paris} (reference ANR-10-LABX-63) part of the Idex SUPER, and received financial state aid managed by the Agence Nationale de la Recherche, as part of the programme Investissements d'avenir under the reference ANR-11-IDEX-0004-02. This research was supported by the DFG cluster of excellence ``\href{www.universe-cluster.de}{Origin and Structure of the Universe}''.}

\section*{References}
\bibliography{vietnam}

\end{document}